# Imaging individual solute atoms at crystalline imperfections in metals

Shyam Katnagallu[1], Leigh T. Stephenson[1], Isabelle Mouton[1, ♣], Christoph Freysoldt[2], Aparna P.A. Subramanyam[3], Jan Jenke[3], Alvin N. Ladines[3], Steffen Neumeier[4], Thomas Hammerschmidt[3], Ralf Drautz[3], Jörg Neugebauer[2], François Vurpillot[5], Dierk Raabe[1], Baptiste Gault[1,6,*]

**Affiliations:**

[1]Department of Microstructure Physics and Alloy Design, Max-Planck-Institut für Eisenforschung GmbH, Düsseldorf, Germany.

[2]Department of Computational Materials Design, Max-Planck-Institut für Eisenforschung GmbH, Düsseldorf, Germany.

[3]ICAMS, Ruhr-Universität Bochum, D-44801 Bochum, Germany.

[4]Materials Science and Engineering, Institute 1, Friedrich-Alexander-Universität Erlangen-Nürnberg, Erlangen, Germany

[5] Normandie Univ, UNIROUEN, INSA Rouen, CNRS, GPM, 76000 Rouen, France

[6]Department of Materials, Imperial College London, Kensington, London, SW7 2AZ, UK

* b.gault@mpie.de, b.gault@imperial.ac.uk.

♣ now at DEN-Service de Recherches de Métallurgie Appliquée, CEA, Gif-sur-Yvette, France.

**Abstract:**
Directly imaging all atoms constituting a material and, maybe more importantly, crystalline defects that dictate materials' properties, remains a formidable challenge. Here, we propose a new approach to chemistry-sensitive field-ion microscopy (FIM) combining contrast interpretation from density-functional theory (DFT) and elemental identification enabled by time-of-flight mass-spectrometry and data mining. Analytical-FIM has true atomic resolution and we demonstrate how the technique can reveal the presence of individual solute atoms at specific positions in the microstructure. The performance of this new technique is showcased in revealing individual Re atoms at crystalline defects formed in Ni during creep deformation. The atomistic details offered by A-FIM allowed us to directly compare our results with simulations, and to tackle a long-standing question of how Re extends lifetime of Ni-based superalloys in service at high-temperature.

Single atom analysis of the structure and composition of matter is of the utmost importance since materials are nowadays designed at this scale. This applies particularly to the quantification of solute decoration at crystalline defects which determine many properties of advanced materials. Valiant efforts have been reported by transmission electron microscopy (TEM) combined with X-ray-based or electron-energy-loss spectroscopy [1]. The inherently three-dimensional nature of dislocations and grain boundaries limits the accuracy of techniques integrating signal through the thickness of a specimen, like TEM. Modern approaches, involving intensive computation, have come close to true atomistic reconstruction through tomography or focus-series [2–4] on model material systems. High-resolution TEM (HR-TEM) performed on specimens prepared for subsequent APT analysis is close to providing the necessary structural and compositional information [5]. Yet, again, TEM provides a two-dimensional, projected image, and the accuracy of APT is limited by trajectory aberrations, precluding direct matching of atomic positions and elemental identity.

Field-ion microscopy (FIM) was the first technique allowing to image individual atoms [6] [7]. It relies on the ionization of an imaging gas, usually a rare gas, caused by an intense electrostatic field generated at the surface of a specimen shaped as a sharp needle. Electrons from the field-ionized gas atoms tunnel through vacuum into the surface, and the ions are projected onto an assembly made of micro-channel plates and a phosphor screen. FIM has brought valuable insights into the structure of crystalline defects, e.g. dislocations and grain boundaries [8,9], in pure metals. The nature of the contrast in FIM has been under debate [10,11], in particular when atoms from different species are involved. The brightness of an imaged atom is expected to be related to its elemental identity as well as to the local topology of the surface, and distinguishing the contribution between these two aspects has so far not been achievable. FIM predates APT, but its analytical capacity enabled APT to rise in importance, despite a significant loss in spatial resolution, whereas FIM is barely used nowadays.

Here, we introduce a dual approach to turn FIM into an analytical single-atom microscopy technique with true atomic resolution in three-dimensions. For this purpose, we use Tersoff-Hamman-type image simulations via density-functional theory (DFT), to provide elemental contrast interpretation of the signal produced by the image gas. This is an interpretation technique known from scanning tunneling microscopy (STM). In parallel, we combine time-of-flight mass-spectrometry (*tof-ms*) and advanced data-filtering techniques to associate an imaged position and an elemental identity. The *tof-ms* enables elemental sensitivity to complement the intrinsically unparalleled spatial resolution of FIM. Realizing these two independent sets of evidence in a single experiment allows us to render FIM chemistry-sensitive, thus revealing position and type of individual atoms.

We showcase the strength of the method on a long-standing problem in alloys designed for high-temperature applications: how Re interacts with crystalline defects to help extend the creep lifetime of Ni-based superalloys in jet engines, enabling higher operating temperatures and fuel efficiency [12,13]. The alloys are hardened by coherent $L1_2$ ordered γ' precipitates formed in a disordered face-centered cubic γ matrix. The third generation of superalloys contains only 5–6 wt.% Re (approx. 2 at.%) which increases creep life by a factor of almost two [14–16]. Re partitions strongly to γ [17–19], but experimental results on its distribution have been conflicting [15,20]. We selected a model alloy containing 2 at.% Re crept under a constant applied compressive stress of 20 MPa at 1050 °C until it reached 5.5% plastic strain after 14.3 hours (see methods). Needle-shaped specimens were subsequently prepared (see Suppl. Material) and were initially characterized by APT, see Fig. 1 (a). The measured Re composition is 1.9±.05at.%, i.e. very close to the nominal composition. No tendency for Re clustering was found in the histogram of Re first nearest-neighbor distances that rather matches a random distribution (Fig. 1(b)), in agreement with previous reports [20].

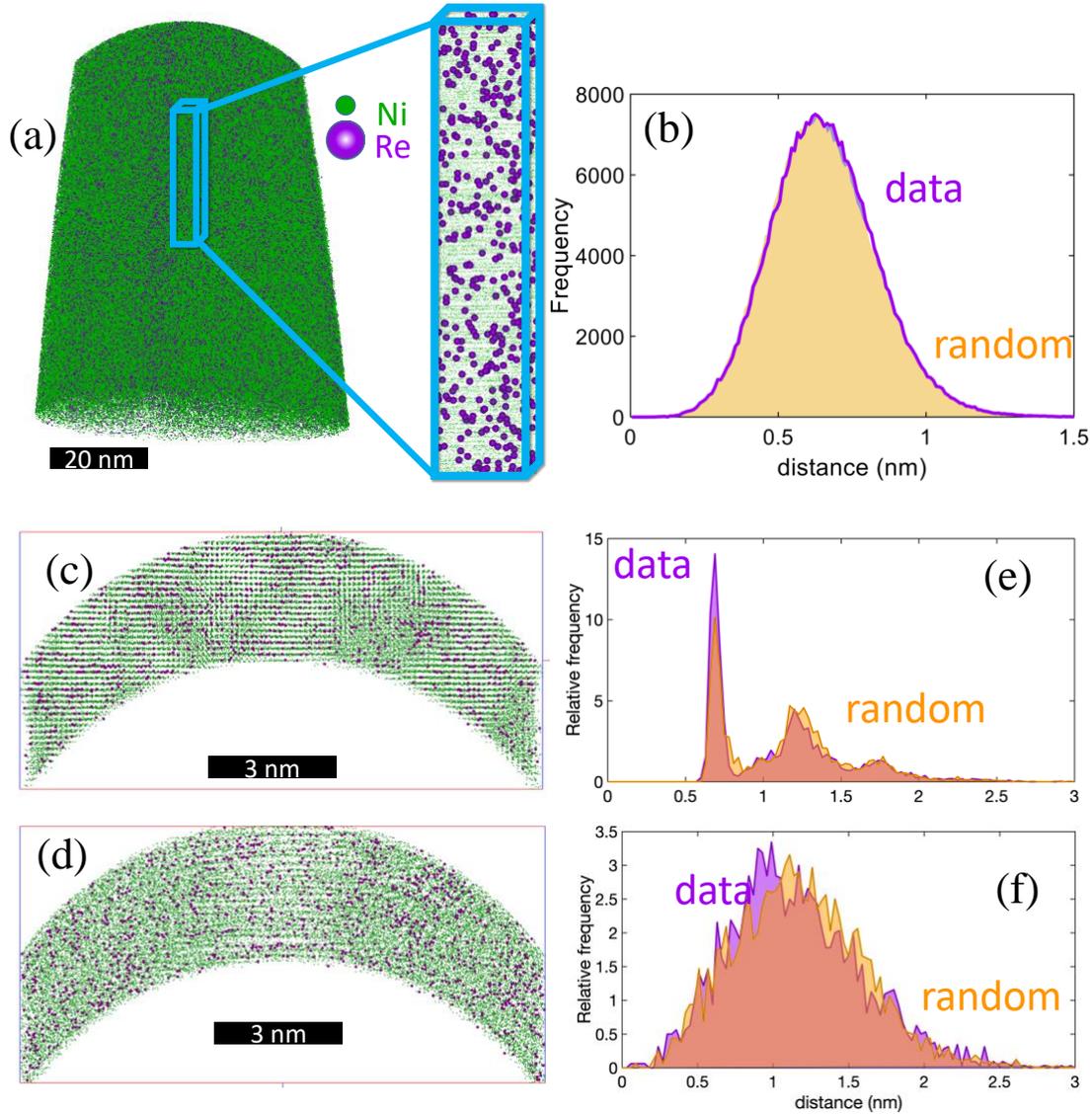

*Figure 1: (a) typical reconstruction from an atom probe analysis of the deformed Ni2Re alloy, with a close-up on the (002) atomic planes imaged within the dataset, and (b) corresponding histogram of the distance between a Re atom and its first Re nearest-neighbor. (c) Simulated deformed pure Ni, in which Re atoms were introduced, to be used as input into APT image simulations. (d) Tomographic reconstruction calculated for the simulated data. (e)–(f) First Re nearest-neighbor analysis for the input and reconstructed datasets.*

To best mimic the structure of a deformed sample, we used the Large-scale Atomic/Molecular Massively Parallel Simulator (LAMMPS) [21] software package to perform molecular dynamic simulations of pure Ni highly deformed under uniaxial compression. The resulting atomic positions were used an input for image simulations of APT. Using dislocation analysis (DXA) [22] in OVITO [23], all dislocations were identified in the input file. About 2% of Re atoms were randomly placed Ni sites, and a 20 at.% concentration of Re (20%) were positioned on stacking faults (more details are in Suppl. Material). Using this approach, a local enrichment of Re close to dislocations is generated. A subset of the original input data is shown in Fig 1(c). This virtual specimen was field evaporated using the Robin-Rolland Model, described in retail in Ref. [24] Because the model is meshless, it is particularly well-suited to model field evaporation of non-regular distribution of atoms, i.e. at or near defects. The evaporation fields of Re atoms was set to 1.3 the evaporation field of Ni atoms, and following field evaporation, the data was reconstructed [25]. A section of the resulting three-dimensional reconstruction is displayed in Fig. 1(d). The distance to the first Re nearest-neighbor performed for the input data and reconstructed data are plotted in Fig. 1(e) and (f) respectively. The former shows a clear peak at approx. 0.23nm corresponding to the agglomerated Re at the defects. The latter, however, is conform to a random distribution, because of the severe trajectory aberrations arising from the close proximity of Re atoms that lead to very sharp protrusions at the specimen's surface during field evaporation and hence to highly divergent trajectories [26]. To avoid this blurring of the atomic position that hinders identification of the segregation, the atomic positions should be imaged prior to field evaporation.

FIM was therefore performed, here with Ne as imaging gas at a pressure of $10^{-7}$ mbar in the same instrument, using the position-sensitive, time-resolved detector typically used in APT. A digitally recalculated field-ion micrograph is shown in Figure 2A, along with a close-up on a (022) terrace displayed in Figure 2B. An edge dislocation can be seen emerging from the terrace, highlighted, in the close-up by the yellow lines. The core of the dislocation is decorated by two brightly imaging atoms. The field ionization process that leads to the formation of the image occurs mainly in an 15 pm-thick shell 400–1000pm above the surface atoms [6]. Noting that this behavior is formally equivalent to the tip in STM allows us to employ a particularly powerful simulation approach known as the Tersoff-Hamann approximation [27]. According to this concept, the electron transfer probability in a 3D scattering theory is proportional to the local DOS at the position of the tip. We performed DFT calculations on steps at (110) facets in Ni running along <100> lines, where individual atoms are experimentally resolved and the probability to transfer an electron from the Ne imaging gas to the surface can be accurately simulated [28]. Here, we included a strong electric field in the calculation to mimic experimental conditions, and evaluated the critical height from the laterally averaged potential. The local DOS was calculated in a plane above a stepped surface, with three step atoms in the view, first for three Ni atoms, then a Re atom with a Ni atom on either side, as depicted in Figure 2C, under an electric field of 50 V.nm$^{-1}$. The Ni/Re contrast is dominated by the tunneling of spin majority electrons. The DOS of bulk Ni has a fully occupied *d*-band in the spin majority channel, so in the absence of field no additional electrons can tunnel. Re has d-band states above the Fermi level for both spins. As the empty spin-majority Re *d*-states cannot couple to Ni states, the electrons remain very localized. In the spin-minority channel, the wave-function is much more delocalized over Re and Ni states. The picture for spin-minority only in the Re case is very similar to the case of pure Ni. Re is hence expected to appear brighter, while the neighboring Ni-atoms should appear dimmer.

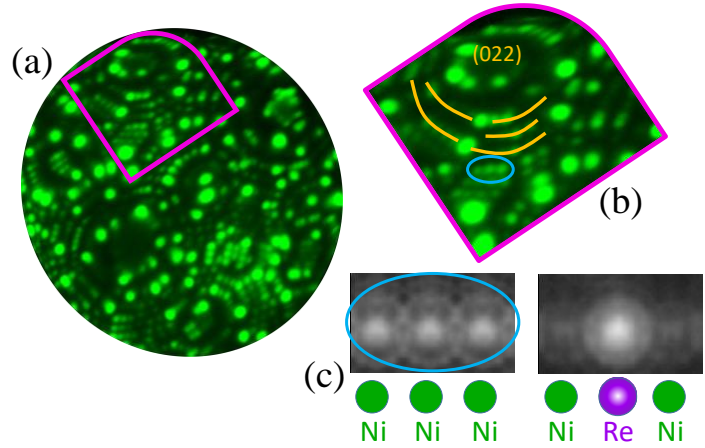

*Figure 2: (a) field ion micrograph and (b) close-up on the (022) planes indicated by the pink box. The presence of a dislocation is highlighted with the light orange lines, as an additional half-plane appears on the right side of the pole. (c) DFT-simulated contrast from the presence of a Re atom in the steps at (110) facets running along <100>. This terrace is indicated by the blue ellipse in (b).*

To obtain further confirmation, we performed *tof-ms* during FIM on the specimen imaged in Figure 3A. Historically, the implementation of a mass spectrometer on a field ion microscope is what became the atom probe, that later led to APT. However, APT is performed under ultra-high vacuum conditions. Here, we maintained an imaging gas pressure of approx. $10^{-7}$ mbar, allowing to map the specimen surface at atomic resolution via imaging gas field ionization. Field evaporation of the surface atoms is triggered by high-voltage pulses superimposed to the DC voltage. We estimate that the rate of imaging gas ionization is three orders of magnitude higher than the rate of field evaporation, i.e. the atom is imaged thousands of times prior to leaving the surface. The mass spectrum from the analysis is shown in dark blue in Figure 3B. The high level of background hinders direct identification of specific mass-to-charge peaks. However, the field desorption and ionization of an adsorbed image-gas atom and the field evaporation of a surface atom are highly likely to be concomitant and hence be emitted by a single pulse making them temporally and spatially correlated [29]. We therefore developed spatial and temporal data filtering, i.e. selecting ions detected on multiple events [30] and arriving within only 2 mm of each other on the detector. The corresponding spectra are shown in blue and light blue respectively in Figure 3B. The associated enhancement of the signal-to-background ratio allows to unmistakably identify the brightly imaging atoms as Re, thereby making FIM truly analytical.

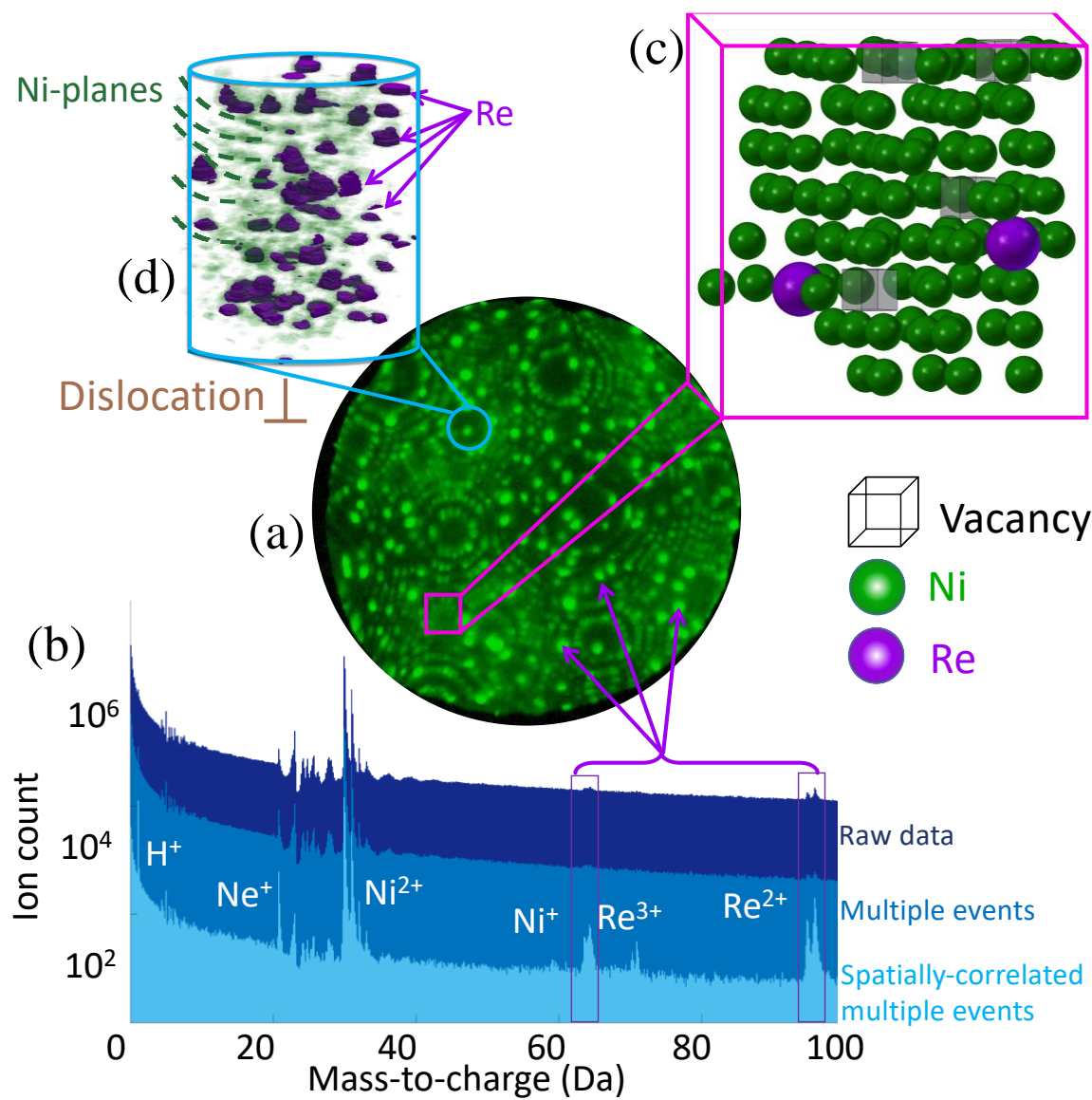

***Figure 3:*** *(**a**) Field ion micrograph. (**b**) Mass spectrum obtained before filtering (in dark blue), after filtering for multiple hits (blue) and after application of the spatial filtering (light blue), allowing to clearly distinguish peaks pertaining to Ni and Re. (**c**) Three-dimensional reconstruction of the atomic positions on the set of (204) planes that do not contain a dislocation. Individual vacancies are imaged and displayed as translucent grey cubes. (**d**) Image stack revealing the local distribution of Re, where a dislocation is intercepting the (204) pole; the faint green lines are traces of the successive Ni-planes that are field evaporated and indicated by the yellow dashed line; the purple isosurface delineates regions of very bright contrast and highlights the Re atoms. Full atomic reconstruction is here achieved in the plane (see Fig. 1) but not in 3D due to a slight resolution loss at the defect.*

Analytical-FIM is much less sensitive to the trajectory aberrations that limit APT's spatial resolution. The atomic resolution of FIM allows to build a fully atomically-resolved three-dimensional reconstruction using the image processing protocols introduced by Katnagallu et al. [31], as shown in Figure 3C. Ni and Re atoms are respectively green and purple spheres, while vacancies are represented as translucent dark grey cubes. Vacancies and Re atoms often appear in close proximity, which could be related to slight attractive interactions discussed by Schuwalow et al. [32], however the statistics here remains limited. In Figure 3D, we produced a stack of images around the dislocation recorded as the specimen gets field evaporated, indicating how Re is distributed in three-dimensions. A slight decrease in spatial resolution and contrast, due to the combined presence of the defect and brightly imaging Re atoms, does not allow for full atomic reconstruction in 3D at this specific position, yet the distribution of individual Re atoms and the trace of each evaporating plane are clear and also atomically resolved in each 2D view (Fig. 1). This new simulation-enhanced approach to analytical-FIM provides the first experimental proof of individual Re atoms segregating to the core of edge dislocations. With this novel single-atom microscopy approach, the composition of Re atoms to two dislocations could be determined to be on average 7.5± 1.8 at. % Re segregated over 20 (2 4 0) planes.

To understand the driving force for Re to segregate, we used atomistic simulations. Modelling segregation of solutes to dislocations is challenging owing to the antipodal requirements of an accurate description of the bond chemistry and large-scale simulations of the long-range strain fields of edge dislocations. The possibility of specific interactions between crystalline defects and Re was investigated by DFT, yet, the results were contradictory [32–35]. We simulated the interaction of Re with an edge dislocation by a combination of tight-binding (TB) and analytic bond-order potentials (BOP) [36] using a new parameterization for the Ni-Re system (see Methods). In parallel, we use DFT to determine the binding energy of Re to a (111) stacking fault in Ni (see Methods). The TB/BOP model provides a transparent description of bond chemistry [37] suitable for large-scale simulations [38], while DFT provides accurate energetics. The binding energy of a Re atom positioned in the supercell either at the stacking fault, at the partial or in the matrix is obtained from the energy of relaxed supercells. We report the binding energy relative to the absolute value of the binding energy of Re at a pure stacking fault derived from the BOP. The tensile part of the core is approximately twice more attractive than the stacking fault. The second moments of the electronic density of states of the Re atom at different positions can be compared to that of the bulk, $\mu_2^{bulk}/\mu_2$, which provides an estimate of the local volume available to the atom. In this position, the Re atom has the largest local volume available, as shown in Figure 4(b). The compressive part of the partial dislocation core is considerably repulsive for Re. The tensile region of the planar fault is also attractive, in agreement with DFT, and it is relatively more attractive than the compressive region. Segregation of Re to full or split edge dislocations results in a reduction of the system's energy and is hence highly favorable. To impose a drag force [39] capable of slowing dislocation climb [33], the diffusivity of Re should be comparable to the climb velocity of dislocations. Our estimations for 1050°C at a load of 20 MPa (see Suppl. Material) indicate that, on average, Re can keep up and hence impose drag as they move through the material during creep at high temperature.

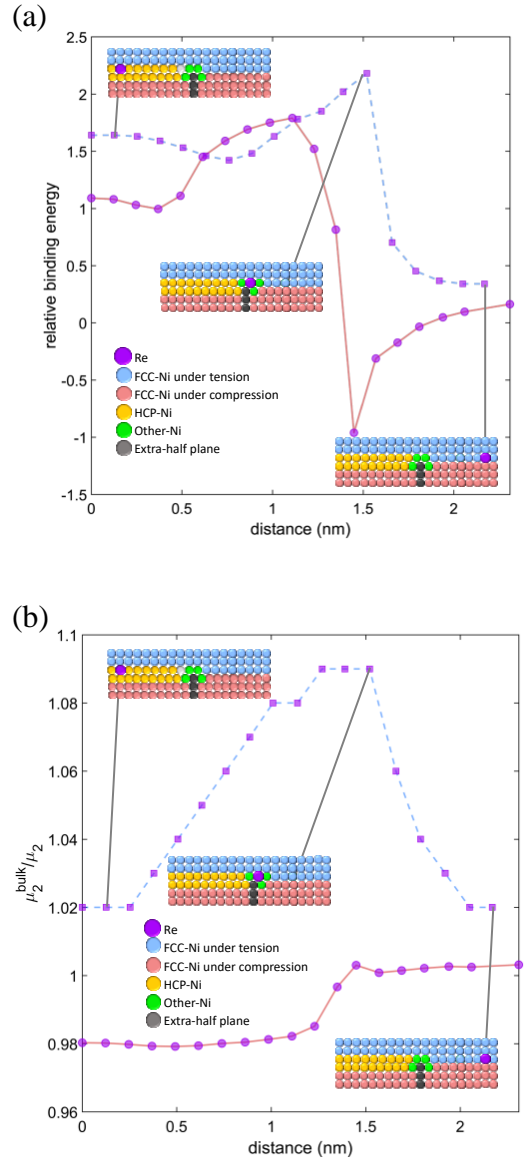

*Figure 4: (a) Relative binding energy of Re at a stacking fault plotted as a function of the distance to the centre of the stacking fault in the tensile (squares, light blue line) and compressive layers (disks, light red line). The atomic models of the corresponding specific configurations are included inset. (b) Relative atomic second moments of the electronic density of states from the BOP plotted as a function of the distance to the centre of the stacking fault in the tensile (squares, light blue line) and compressive layers (disks, light red line). The atomic models of specific configurations are included inset.*

Rationalizing empirical know-how at the atomic scale lays the grounds for smarter materials design to achieve enhanced performances and keeps inspiring continued improvements in atomic-scale microscopy and microanalysis. Here, we introduced an approach to make FIM truly sensitive to single atoms, with atomic spatial resolving power resolving and chemistry-sensitivity. Analytical-FIM allowed us to investigate the segregation of individual solute atoms to crystalline defects in a high-temperature creep-deformed Ni-Re alloy. Complemented by DFT-based contrast interpretation and *tof-ms*, the new approach to analytical-FIM provides clear evidence of the segregation of individual Re atoms to dislocations, in contrast to a conventional APT analysis. Analytical-FIM proves to be superior to APT in that respect since the atomic positions are revealed prior to field evaporation by a statistical distribution of thousands of image-gas ions, the trajectory of which are less affected by trajectory aberrations. Analytical-FIM opens new opportunities to deepen the understanding of how specific solutes interact with crystalline imperfections, including dislocations, vacancies and potentially, in the future, grain boundaries.

**Acknowledgments:** The authors are indebted to Dr. Hamad ur Rehman for performing the creep tests. BG is grateful for many discussions with Drs Paraskevas Kontis, Surendra Kumar Makineni, Junyang He and Xiaoxiang Wu. SK, LTS, IM and BG are grateful to Dr Richard Forbes for discussions. The authors are grateful to U. Tezins and A. Sturm for their technical support of the APT and FIB facilities at the Max-Planck-Institut für Eisenforschung.

**Funding:** SK acknowledges the International Max-Planck Research School SurMat for funding. BG, LTS, SK and IM acknowledge funding from the MPG through the Laplace project. SN, BG, APAS, RD, TH acknowledge financial support from the German Research Foundation (DFG) through projects A4, B3 and C1 of the collaborative research centre SFB/TR 103. LTS and BG acknowledge financial support from the ERC-CoG SHINE – 771602.

**Author contributions:** SK, LTS, IM developed the ideas for analytical-FIM, the data processing routines, with support and input from BG. BG conceived the project on the NiRe. SN provided the samples and the discussion on the diffusivity of Re in Ni. SK performed the APT, FIM and analytical-FIM experiments. CF and JN introduced the idea of the T-H calculations and CF implemented code and performed the DFT calculations. APAS, JJ, ANL, RD, TH developed the TB/BOP and performed the DFT/TB calculations. FV performed the field evaporation simulations based on input from SK, Results were discussed with all authors. BG drafted the manuscript with input from DR, SK, TH, FV. All authors were given a chance to comment.

**Supplemental Material**

Material

Details of the manufacturing and heat treatment of the single crystal rod of the Ni with 2 at.% Re can be found elsewhere [39]. A cylindrical specimen was then subjected to uniaxial compression in a pneumatic compression creep testing machine (Insitron 405) at a constant applied stress of 20 MPa at 1050°C until a plastic strain of 5.5% was reached within 14.3 hrs.

FIM and Atom probe

Specimens for FIM and APT were prepared on a FEI Helios PFIB. The usual protocol by Thompson et al. was followed\cite{Thompson2007b} in a Xe-plasma FIB, with milling current at 30KeV between 0.46 nA~24 pA and a final cleaning at 5 keV and 24 pA to remove a possibly damaged regions. The APT experiments were conducted on a Cameca LEAP 5000 XS.

Ultra-high purity Ne gas was supplied directly into the analysis chamber via a manually controlled leak valve. The gas pressure was adjusted so as to maintain a pressure in the range of $1.3\pm0.5\ 10^{-7}$ mbar. Time-of-flight FIM was performed on the same instrument, with high-voltage pulses and a pulse fraction of 20%. A similar approach with laser pulses was recently discussed in Ref. [40].

The APT data reconstruction and post-processing was done in the commercial software IVAS 3.8.2. FIM data extraction in the form of a .epos file came from the same software package, and further processing was done via in-house built routines in MATLAB. A first pass filter is applied following the idea that ions generated at a higher electric field during the pulse will tend to arrive on higher order of hit multiplicity [30]. These correlated field-evaporative events are also known to be spatially correlated [29], and hence a second pass filter is applied whereby only ions arriving within a radius of 2mm of each other are considered.

DFT for FIM contrast interpretation

We used DFT in the local-density approximation using the projector-augmented wave approach [41] as implemented in the SPHInX code [42]. The surfaces were modelled in the repeated slab approach, with a thickness of six layers along the (011) normal (approx. 1.4 nm) and a vacuum separation of 1.5 nm. The slabs were subject to a field of 50 V/nm on one side using the generalized dipole correction. Owing to the difficulty of convergence in DFT for high fields, a field of 50 V/nm was chosen. Test calculations show that variations in the field strength by over 10% do not qualitatively change results. The imaging plane in the vacuum for the DOS calculation was chosen such that the average electrostatic potential on that plane corresponds to 20.8eV above the Fermi energy, i.e. the ionization threshold for Ne. There is some lateral variation of the potential of the order of a few tenths of an eV. Unoccupied states up to 0.5 eV above the Fermi level (sharp cut) were included; the occupied states are cut by an artificial Fermi function ($e_{kt}$=0.1 eV).

Atomistic simulations for Re segregation in Ni and superalloys

For Re in superalloys, recent work showed indication of Re segregated to interfacial dislocations in a crept state [52], without allowing to quantitatively resolve its distribution at the atomic scale and using a potential that is known to have deficiencies [53]. indicating only a low binding energy of Re to vacancies unlikely to significantly slow down their diffusion [32]. Liu et al. [35] used

DFT combined with a lattice Green-function boundary condition treatment to derive the binding energy of Re to a split dislocation in Ni. These results suggested that Re should segregate to the stacking fault rather than to the adjacent core of the partial dislocations.

The spin-polarized DFT calculations of Re segregation to a pure stacking fault were performed using the VASP software package [43] with the Perdew-Burke-Ernzerhof [44] exchange-correlation functional, projector augmented-wave (PAW) [41] pseudopotentials and a plane-wave basis with 400 eV cut-off energy. For the three supercells with 4, 8 and 16 atoms per (111) stacking-fault layer and 11 layers in [111] we used Monkhorst-Pack [45] k-point meshes of 10x10x2, 5x10x2 and 5x5x2, respectively. Using an electronic convergence of $10^{-6}$eV, the atoms were relaxed until a maximum force of less than $10^{-2}$ eV/Å. Our DFT calculations predict an attractive interaction of Re with the two symmetry-equivalent layers of the stacking fault with an energy of 120 meV for a concentration of 1/4 Re atoms per stacking-fault layer and 100 meV in the dilute limit of 1/32, in agreement with literature [48].

The TB/BOP calculations are performed with the BOPfox [38] using a non-magnetic $d$-valent Hamiltonian in analytic BOP [36] and *k*-space TB, respectively. The parameters for the Ni-Re system are obtained by down folding the DFT eigenspectrum to a tight-binding minimal basis [46] followed by an optimization with an additional pairwise repulsive term to reproduce properties of Ni-Re bulk phases. The TB/BOP model used here gives a stacking fault energy in Ni $\gamma_{ISF}$ approx. 120 mJ/m$^2$, in good agreement with literature [47,48].

For the TB/BOP simulation an a/2[110] edge dislocation is introduced into a Ni supercell with 3200 atoms that, upon relaxation, dissociates into two a/6[112] partials separated by a (111) stacking fault. Simulations are performed in slab calculations with periodic boundary conditions in the stacking fault plane and 12 atomic layers above and below the fault plane. The slab is repeated two times along the dislocation line of the partials, equal to a distance of 8.63Å, distance between the Re atom and its periodic image. A Re atom is then introduce at different positions to investigate the energetics of segregation to the dislocation partials and the stacking fault in Ni. After atomic relaxation for each Re position, the binding energy of Re to the dislocation is determined with respect to the formation energy of the dislocation and the solution energy of Re as: $E_B^{Re-dislocation} = E_F^{Re,dislocation} + E_F^{no\ Re,\ no\ dislocation} - E_F^{Re,no\ dislocation} - E_F^{no\ Re,dislocation}$ with a value of zero in the limit of no Re-dislocation interaction and negative values for attractive interaction. The non-vanishing value of $E_B^{Re-dislocation}$ at the largest Re-dislocation distance in Figure 4, indicates a small finite-size effect of our supercell calculations of few 10meV.

Composition measurement by FIM
The Re composition in FIM was estimated based on counting the number of Re atoms in an image through thresholding and identification of individual atoms as outlined in Ref. (26), and comparing to the number of atoms within a shell of the material at the surface of the specimen, the area of which was determined by measuring the radius of the specimen via ring counting [49]. The Re composition is 1.94 at.% close to the composition of the alloy measured by APT. However, the error, difficult to estimate, is likely large due to the relatively low statistics.

In Figure 3(d), eight terraces were evaporated in a sequence of FIM images over which 103 atoms evaporated. As the original images were relatively noisy, an additional image processing routine was employed. Similar to averaging method in Ref. [31], the images were averaged to improve the signal-to-noise ratio and a logarithmic filter on the intensities was applied owing to the severe contrast between Re and Ni atoms. The atom positions were extracted using the atom detection algorithm described in Ref. [31]. However, for an efficient plane classification in the images required manual input due to the retention of Re atoms at the surface.

Density Functional Theory calculations for FIM contrast interpretation
A (023) surface of fcc Ni was simulated. The experimental result shown in Figure 1A also shows such configurations (see ellipse in the close up). A (023) surface is made of (011) terraces with steps running along <010>, such that the atoms on the step are separated by the cubic lattice constant $a$. For the surface whose step atom's separation is $a/\sqrt{2}$ along <011>, the resolution is insufficient to resolve the individual atoms well. Such a strong field not only influences the surface relaxation, but also leads to a partial depletion of near-surface states. The explicit inclusion of the electric field was therefore deemed necessary to reflect the experimental situation. This is in contrast to conventional STM simulations, where the electric field is commonly neglected because it is 1-2 orders of magnitude smaller (in STM, a typical bias is on the order of 1V at a tunneling distance on the order of 1 nm). Indeed, Tersoff-Hamann simulations in the absence of field show only marginal contrast (1:1.2) along the step for the pure Ni case at the height relevant for FIM. A careful examination also reveals a "corona" around the Ni atoms, which might be an artefact of insufficient k-point sampling or finite slab thickness. Overall, these calculations provide a good qualitative match with what is seen experimentally, i.e. imaging simulations conducted along the model approximations of the Tersoff-Hamann model render the FIM chemistry sensitive.

Image simulations for atom probe
In order to better understand the experimental results, complementary FIM and APT simulations were employed to obtain detailed information about the artefacts inherent to both techniques, which are subject to different image distortions mechanisms. In FIM the contrast arises from the atomic-scale topography of the surface, but also from the tunneling probability that vary from surface atom to surface atom, as we demonstrated for the first time herein. In APT, ions are field evaporated from their atomic locations at the surface, and the ions are projected onto the detector. The corrugation of the surface, resulting from the atomic neighborhood of each atom, may cause significant fluctuations of the direction of the electrostatic field from place to place. In alloys, atoms can evaporate at different field strengths that depend strongly of their elemental natures. Classical theories predict an evaporation field for Ni of 35 V/nm, and 45-48 V/nm for Re. [50]. In Subsequently, the ratio of evaporation fields will be considered to be 1.3. At a constant field evaporation rate, Re atoms will tend to preferentially be in retention at the specimen's surface, in order to enhance the local electric field sufficiently to cause field evaporation.

To best mimic the deformed structure, we used the Large-scale Atomic/Molecular Massively Parallel Simulator (LAMMPS) [21] software package to perform molecular dynamic simulations of a Ni sample highly deformed by uniaxial compression. The resulting atomic positions were used an input for image simulations of FIM and APT. The deformed volume was replicated using periodic boundary conditions, and then reduced into a tip with an initial radius of 12 nm. The input contains 1.4 millions of atoms. Using dislocation analysis (DXA) [22] in OVITO [23], all dislocations were identified in the input file. The different crystalline environments were used to identify local hexagonal-close packed (HCP) and face-centered cubic (FCC) structures corresponding to stacking faults and the metallic matrix respectively. About 2% of Re atoms were randomly placed on non-HCP Ni sites, and a high concentration of Re (20%) were positioned on HCP Ni sites. Using this approach, a local enrichment of Re close to dislocations is generated.

The sample was field evaporated using the Robin-Rolland Model. The model uses the electrostatic Robin's equation to directly calculate charge distribution over the tip apex conducting surface, without the need for a supporting mesh. Each surface atom is considered as a point charge, which is representative of its evaporation probability. The point charge distribution of atoms at the tip surface gives a direct access to the field distribution at any step of the evaporation process anywhere in the simulation volume. This model is described in retail in Ref. [24] Because the model is meshless, it is particularly well-suited to model field evaporation of non-regular distribution of atoms, i.e. defects. The evaporation fields of Re atoms was set to 1.3 the evaporation field of Ni atoms. About 150000 atoms were field evaporated one by one, and the trajectories of the emitted ions are used to determine the impact positions on a virtual detector. Using the impact position as input into the conventional data reconstruction algorithm [25], a 3D reconstruction of the volume is generated. Atomic planes appear perpendicular to the tip's apex, which denotes that the spatial resolution in depth remains sufficient to obtain layer-by-layer field evaporation, yet the lateral resolution has blurred the lattice positions. This loss of resolution has a dramatic effect on the direct detection of cluster and segregation of Re. A cluster-search algorithm was applied to the initial and reconstructed volumes. Note that a detection efficiency of 80% was set in the reconstructed image to mimic experiments. Using the maximum separation method [51], with a threshold distance for the cluster search of 0.3 nm and a minimum size of clusters of 4 atoms, about 40 clusters are detected in the input volume. However, in the reconstruction after the analysis, only 4 clusters remain.

Dislocation vs. solute mobility

The dislocation velocity at 1050°C and under a load of 20 MPa was compared to the velocity of the diffusing Re atom, in order to see if Re could exert a drag force on a moving dislocation. Ur Rehman et al. [39] reported estimations of the dislocation velocities in the Ni-2Re alloy is approximately 2 nm.s$^{-1}$. Karunaratne et al. [54] reported the diffusion coefficients. The diffusion length $\sqrt{D\,t}$D for a Re atom under the same conditions is approx. 8 nm.s-1. These values are comparable, and it is hence possible that Re imposes a drag on the dislocation. As a comparison, W and Ta diffuse approx. 17 nm and approx. 52 nm in 1s respectively, i.e. much faster than Re, which means that they would not hinder the dislocation movement as strongly as Re, consistent with experimental observations [39].

**Supplementary References**